# SD-WAN Internet Census

**Sergey Gordeychik**

Inception Institute of Artificial Intelligence

Abu-Dhabi, UAE

serg.gordey@gmail.com

**Denis Kolegov**

Tomsk State University

Tomsk, Russia

d.n.kolegov@gmail.com

**Antony Nikolaev**

Tomsk State University

Tomsk, Russia

antoniy.nikolaev@gmail.com

## ABSTRACT

The concept of software-defined wide area network (SD-WAN or SDWAN) is central to modern computer networking, particularly in enterprise networks. By definition, these systems form network perimeter and connect Internet, WAN, extranet, and branches that makes them crucial from a cybersecurity point of view. The goal of this paper is to provide the results of passive and active fingerprinting for SD-WAN systems using a common threat intelligence approach. We explore Internet-based and cloud-based publicly available SD-WAN systems using the well-known "Shodan"[1] and "Censys"[2] search engines and custom developed automation tools and show that most of the SD-WAN systems have known vulnerabilities related to outdated software and insecure configuration.

**Categories and Subject Descriptors**
C.2.4 [Computer-Communication Networks]: Network Architecture and Design; C.2.4 [Computer-Communication Networks]: Distributed Systems;

## KEYWORDS

SDN, SD-WAN, security, cybersecurity, threat intelligence, scanning, fingerprinting, cloud networks, vulnerabilities

## INTRODUCTION

The SD-WAN technology is rapidly becoming very popular in enterprise networks. SD-WAN supports security services, such as firewalls, VPN, DPI, IDPS that makes it attractive targets for attackers. Vendors promise "on-the-fly agility, simplicity, security, and automation" and many other benefits. Unfortunately, in practice, most of SD-WAN solutions are distributed as Linux-based virtual appliances or cloud-centric services employing outdated open source software with known vulnerabilities. All these make SD-WAN low-hanging fruit even for a script kiddie.

## Problem Description

The goal of this paper is to provide the results of passive and active fingerprinting for SD-WAN systems using a common threat intelligence approach and to answer the following questions:

- Is SD-WAN secure at Internet scale?
- What is SD-WAN security level in a general sense at the present time?
- Is there any evidence that native centralized management within SD-WAN is effective in practice from a security point of view?
- How long does it take to patch vulnerabilities, apply security updates to the SD-WAN systems deployed on the Internet, and update images (AMI) in the AWS Marketplace?

## Our Approach

The common approach to SD-WAN system enumeration is based on obtaining complete and exhaustive information about a specific SD-WAN solution from each related vendor that is in scope. That information helps us find publicly available SD-WAN devices with well-known search engines such as Shodan and Censys. It is also useful in developing tools for fingerprinting. Very generally, this approach includes the following steps:

1. Looking for major SD-WAN vendors such as Versa, Cisco, Citrix, and many others.
2. Obtaining information about the primary SD-WAN solutions from the vendors for further usage in enumeration and fingerprinting.
3. SD-WAN mass enumeration using passive fingerprinting with Shodan and Censys search engines.
4. Reducing false positive results, adjusting filters, and query correction.
5. Classification of search queries according to confidence level.
6. Product version leakage detection using manual analysis of HTTP response headers, HTML, JavaScript code, etc.
7. Product version extraction by active fingerprinting with Nmap and its NSE engine.

We explored the following SD-WAN products deployed and accessible on the Internet:

1. VMWare NSX SD-WAN: VeloCloud Network Orchestrator
2. TELoIP VINO SD-WAN: TELoIP Orchestrator API
3. Fatpipe Symphony SD-WAN: WARP
4. Cisco (Viptela) SD-WAN: vManage
5. Versa Networks: Versa Analytics, Versa FlexVNF, Versa Director
6. Viprinet: Virtual VPN Hub
7. Riverbed SD-WAN: Riverbed SteelConnect, Riverbed SteelHead
8. Citrix NetScaler SD-WAN: Citrix SD-WAN Center, Citrix NetScaler SD-WAN VPX
9. Silver Peak SD-WAN: Silver Peak Unity Orchestrator, Silver Peak Unity EdgeConnect
10. Nuage Networks SD-WAN: Nuage SD-WAN Portal, VNS Portal
11. Talari SD-WAN: Talari Appliance

12. Huawei SD-WAN: Agile Controller
13. SONUS SD-WAN: SBC Edge, SBC Management Application
14. Arista Networks EOS: Arista Switch
15. Glueware SD-WAN: Gluware Control
16. Barracuda Networks: Barracuda CloudGen Firewall

It should be noted that we interacted with SD-WAN systems using their management plane interfaces, but not WAN interfaces or any other interfaces within the data plane. Whereas previous management interfaces were generally exposed in out-of-band or in isolated networks. In SD-WAN, these interfaces are used to communicate with orchestrators, VNF managers and other elements within NFVI. Thus, in general, the accessibility of management interface on the Internet indicates the presence of CWE-749 weakness "Exposed Dangerous Method or Function".

## Contribution

The main contributions of this paper are the following:
1. We conduct the first known SD-WAN security research at Internet scale.
2. We develop search engine queries and fingerprinting tools for SD-WAN systems.
3. We evaluate the security level for SD-WAN products based on gathered statistics on software versions and known vulnerabilities.
4. We detect previously unknown product version disclosure sources.
5. We show that SD-WAN management interfaces are accessible on the Internet despite that they are critical entry points designed for management and orchestration.

# SD-WAN IDENTIFICATION AND FINGERPRINTING

## Information Gathering

The exploration of SD-WAN solutions and their interfaces (Web UI, REST, SSH, SNMP, etc.) includes a sequence of steps, and the first one is information gathering. Basically, we need to define the list of the most popular vendors engaged in SD-WAN research – any search engine can be leveraged for this purpose by making some rather trivial queries like "SD-WAN solutions" or "List of SD-WAN vendors".

After that, we have the list of SD-WAN vendors currently active. For each of them, we find available solutions including hardware models, software names, and any other additional details like information about web controller, operating system, web interfaces and so on. This information will be used in the subsequent phases.

Often, all the necessary information about an SD-WAN solution can be found on the vendor's website including possible web interface demonstrations exposing URL patterns, web page titles, unique strings, etc.

# Passive Fingerprinting

## Shodan and Censys Search Filters

After we have obtained the necessary information about SD-WAN solutions of the targeted vendor, we can use Shodan and Censys search engines to find SD-WAN interfaces. The main idea is to use the same titles as product names. We can use the following queries in Shodan:

- title:"Viptela vManage"
- title:"Cisco vManage"
- title:"SBC Management Application"

The equivalent queries for Censys:

- 80.http.get.title: "Viptela vManage"
- 80.http.get.title: "Cisco vManage"
- 80.http.get.title: "Sonus SBC Edge Web Interface"

The other great features of Shodan and Censys are favicon, header, and html hash filters. We can find a good example which can be considered as a reference result, get hash of its favicon or header, and then search for other equal resources with it. The example for Shodan:

- http.favicon.hash:-904700687
- http.favicon.hash:-1352019987

We found SD-WAN solutions that have equal favicon and favicon hash. In this cases, it is better to use *http.html_hash method*. The example of hash search from Censys query based on sha256 hash:

- 80.http.get.body_sha256: 63575152efde5bec3ab2a28a502f7a15de7146e2b0fdce47ab0bb699676fb66f

The other common technique could be based on finding devices with TLS/SSL certificates containing keywords:

- ssl:"O=Viptela Inc"
- ssl: "Viprinet" port:"161"

The equivalent examples for Censys:

- 443.https.tls.certificate.parsed.fingerprint_sha256: "ad4c8962d687837c54a3430e869aadfc359db7fd07d9b0630ec2f355aa7b896a"
- 443.https.tls.certificate.parsed.issuer.common_name: "vmanage"

## Reducing False Positive Results

Sometimes, it is hard to find systems with one query so it is helpful to use two or more subqueries in one search. For example, we can search for VMWare NSX SD-WAN with combined query as following:

- http.favicon.hash:-2062596654 title:"301 Moved Permanently"

The query uses *favicon hash* and *title* filters at the same time. It helps us exclude wrong results. This method can be used with a set of different titles which are used for different products:

- http.favicon.hash:1069145589
- http.favicon.hash:1069145589 title:"SD-WAN*Portal"
- http.favicon.hash:1069145589 title:"Architect"
- http.favicon.hash:1069145589 title:"VNS portal"

To reduce the number of wrong results excluded, search is employed. For example, if we want to get all Nuage Networks products based on favicon hash search filter without "VNS portal" and "Architect", we can use the following:

- http.favicon.hash:1069145589 -title:"VNS portal" -title:"Architect"

## Mask and Query Correction

In some cases, valid queries are not so obvious or do not cover all desired results. For instance, if we only know some parts of a query such as name or vendor and some keywords such as "Log in", "Authorization", "Portal", or similar, we can use query masks (expressed by as asterisks) to match unknown or dynamic parts.

A similar method may rely on every word as different subquery of one filter type. If we want to find all the "Log in" pages for FatPipe Networks services, we can use keywords "Log in" and "Fatpipe" as different subqueries of Shodan "title" filter type. The examples for Shodan are:

- FatPipe WARP | Log in
- FatPipe VPN | Log in
- FatPipe MPVPN | Log in

The example for Censys:

- 80.http.get.title:"FatPipe" AND "Log in"

If we want to exclude all results that have "VPN" and "MPVPN" keywords in page titles of previous query results, we can use the exclusion operator:

- title:"FatPipe" title:"Log in" -title:"VPN" -title:"MPVPN"

## Query Confidence

Now, when we have described the basic ideas, enumeration techniques, employed filters, and search engine features, it is obvious that they sometimes do not work and search results can contain false positive findings. To address this issue, we propose to classify enumeration queries according to confidence level as "Certain", "Firm" or "Tentative". This reflects the reliability of the query that was used to identify a class of SD-WAN systems.

The examples of queries that can be classified as "Certain":

- title:"Flex VNF Web-UI"
- title:"VeloCloud Orchestrator"

- title:"Teloip Orchestrator API"

The examples of queries that can be classified as "Tentative":
- title:"Ecessa"
- title:"128T Networking Platform"

## Active Fingerprinting

The methods described above can be used to identify and enumerate most of SD-WAN systems. However, using only passive methods is not always possible.

First, it can lead to false positives because search engine databases do not always contain the necessary information to accurately recognize a device vendor, family, type, and version. For instance, it is usually hard to distinguish between several products belonging to different product ranges of the same vendor.

Second, search engines only support a restricted set of protocols and network mechanisms. For example, Censys does not support the SNMP protocol, and Shodan does not support the Websocket protocol. During our research, we found an SD-WAN vendor that could be identified by a banner in Websocket messages.

And third, none of the known security oriented search engines can be used to retrieve data using custom flows. For instance, product version can be leaked via JavaScript or CSS files. To retrieve JavaScript or CSS files, an attacker has to get a root web page, parse its <script> sections, then download JavaScript code, and extract the product version from JSON payload. Thus, there is a need to actively identify target devices.

During the research, we identified version disclosure issues for the following products (examples shown in Table 1):
1. Cisco (Viptela) vEdge
2. VeloCloud Network Orchestrator
3. TELoIP Orchestrator API
4. Fatpipe Symphony SD-WAN WARP
5. Versa Analytics
6. Versa FlexVNF
7. Viprinet Virtual VPN
8. Riverbed SteelHead
9. Citrix NetScaler SD-WAN VPX
10. Silver Peak Unity Orchestrator
11. Silver Peak Unity Edge Connect
12. Talari SD-WAN Appliance
13. SONUS SBC Edge
14. Sonus SBC Management Application

To achieve this, we developed a set of Nmap NSE [3] and Python scripts. Nmap and NSE scripts can be used to identify version for most products.

At the same time, certain mechanisms such as SSH Warning Message (/etc/issue) may result in Cisco (Viptela) product version information disclosure. Currently popular and widely known security tools like Shodan, Censys, Nmap NSE, and Masscan cannot be used to detect this behaviour. They analyze SSH banners, versions, or ciphers, but not warning messages because this behavior is very specific and unusual. To address this case, we first get IP-addresses in AWS where OpenSSH is running using the following Shodan query:

- "SSH-2.0-openSSH_7.3" port:22 org:"Amazon.com"

and then use a custom Python script that connects to OpenSSH services on the list and analyzes their warning messages. If the message contains the "viptela" keyword, we add this host to the results.

The developed scripts can be used to refine search engines scanning results. Moreover, Nmap scripts allow detection of vulnerabilities and execution of security analysis for SD-WAN infrastructures inside enterprise networks.

We distinguish between two types of version leakages: direct and indirect. A *direct version leakage* makes it possible to clearly determine the version of the product. For example:

- <h5>9.1.2r142</h5>
- https://e.com/8.1.4.11_66255/php/user_login.php

An *indirect version leakage* cannot be used to identify the exact version of the product, but in some cases it still enables us to get information on software updates. For example:

- {ANALYTICS_PATH : "/analytics/v1.0.0/", …}
- <meta name="application-name" content="web3 v0.15.8" />

The idea here is the following. Suppose the most recent build version is "9.6.5". Also suppose this build fixes a critical security vulnerability and leaks the indirect version "0.3.4". If the explored system leaks the indirect version "0.3.1", then we can derive that this system is vulnerable and outdated.

## SD-WAN VULNERABILITY IDENTIFICATION

In some cases, information obtained using active and passive fingerprinting methods can be used to detect known vulnerabilities in configuration.

For example, the Shodan search engine uses the default community strings "public" and "private" for probing SNMP enabled devices.

Successful connection over SNMP with those SNMP community strings can be classified as CWE-798 weakness "Use of Hard-coded or Default Credentials".

Some of the detected vulnerabilities make it possible to obtain extended information about a device: operating system version, web server version, API methods, etc. Comparison of this information with known weakness lists can be used to identify vulnerabilities in the device such as multiple vulnerabilities in Silver Peak SD-WAN products [4]. The summary statistics on the identified vulnerabilities is shown in the next section.

Table 1. SD WAN versions leakages.

| Vendor | Source | Leakage | Example |
|---|---|---|---|
| Cisco vEdge<br>Cisco vBond<br>Cisco vManage<br>Cisco vSmart | SSH warning message (/etc/issue) | Product version | viptela 17.2.4 |
| VeloCloud Network Orchestrator | HTML | UI version | <link href="/css/vco-ui.3.0.0.1509625568730.common.css"> |
| Teloip Orchestrator API | JSON | API version | {"host":"_v5.02_Teloip Orchestrator API", …} |
| Fatpipe SYMPHONY SD-WAN | HTML | Product version | <h5>9.1.2r142</h5> |
| Versa Analytics | JavaScript | Indirect version | {ANALYTICS_PATH : "/analytics/v1.0.0/", …} |
| Versa FlexVNF | JavaScript | Package version | {"package-name": "versa-flexvnf-20161214-191033-494bf5c-16.1R2", …} |
| Riverbed SteelHead | HTML | Indirect version | <meta name="application-name" content="web3 v0.15.8" /> |
| Citrix NetScaler SD-WAN VPX | HTML | Product version | <link href="/br_ui/rdx/core/css/rdx.css?v=9.3.1.35" rel="stylesheet" type="text/css"/> |
| Silver Peak Unity Orchestrator | URI | Product version | https://e.com/8.3.6.35923/webclient/php/login.html |
| Silver Peak Unity EdgeConnect | URI | Product version | https://e.com/8.1.4.11_66255/php/user_login.php |
| Talari Appliance | URI | Indirect leakage | /css/talari.css?R7_3_QA_P1_D1_06152018 |
| Sonus SD-WAN SBC Edge | HTML | Product version | <link rel="stylesheet" type="text/css" href="/style/6.1.2-471_rel_style.css"> |
| Sonus SD-WAN SBC Management | HTML | Product version | <div id="header_app_name">EMA 6.2 <span class="header_app_name_mode"> |
| Arista vEOS | SNMP | Product version | Arista Networks EOS version 4.15.6M running on an Arista Networks DCS-7150S-24 |
| Glue Networks Gluware Control | Websocket | Product version | <content-bar-item>v3.3.98 (July 10, 2018 6:27am)</content-bar-item> |

# SCANNING RESULTS

Using the provided methodology we conducted an analysis of SD-WAN systems accessible on the Internet. 4935 unique IP-addresses belonging to SD-WAN management interfaces were detected in August 2018 using queries with "Certain" level of confidence (see figures 1 - 3).

According to these results, most of the remotely accessible hosts with SD-WAN components (by unique IP addresses) are located in the North America (1300 hosts), in Europe (456 hosts), and in Asia (313 hosts).

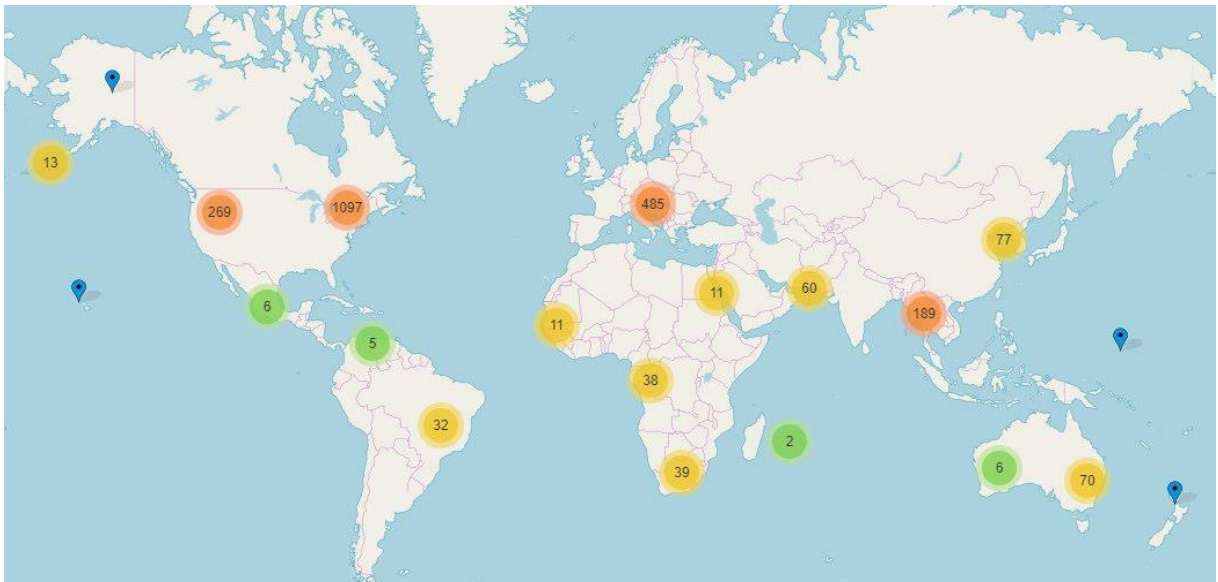

Figure 1. SD-WAN availability by geo IP

The most widespread systems likely belonging to the enterprise segment are from the following vendors: Riverbed (493 hosts), VMWare (452 hosts), Arista (255).

The Table 2 based on SD-WAN systems actual version information shows that most of the SD-WAN deployed on the Internet use outdated software containing known vulnerabilities. Moreover, virtual appliances provided by SD-WAN vendors using cloud services (e.g. AWS Marketplace) are outdated as well. The existence of outdated systems increases the risk of successful attacks against SD-WAN systems.

Consider an example. According to [5] on August 14th, 2018, Arista released information about a denial of service vulnerability where a crafted IP fragment ordering or overlap can allow an attacker to consume more memory than defined in the Linux kernel settings. It is reported that Arista EOS, vEOS, CloudVision Portal, and CloudVision Appliance are affected products. Affected EOS versions are 4.20.X and 4.21.0F. At the same time as we write this AWS Marketplace provides Arista EOS-4.20.5F-BYOL.

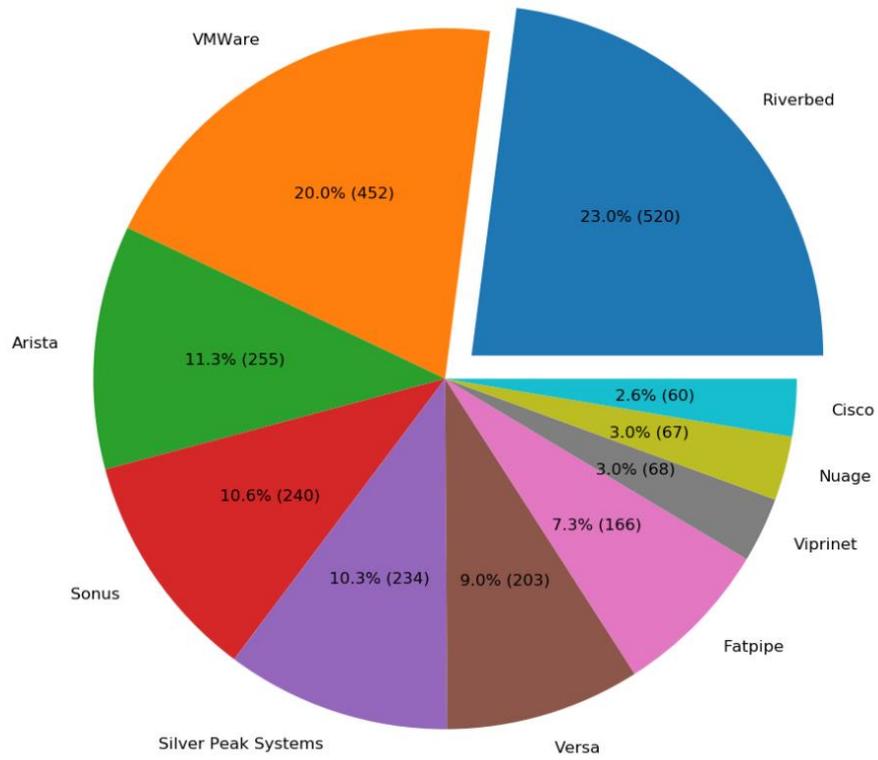

Figure 2. SD-WAN vendors on the Internet.

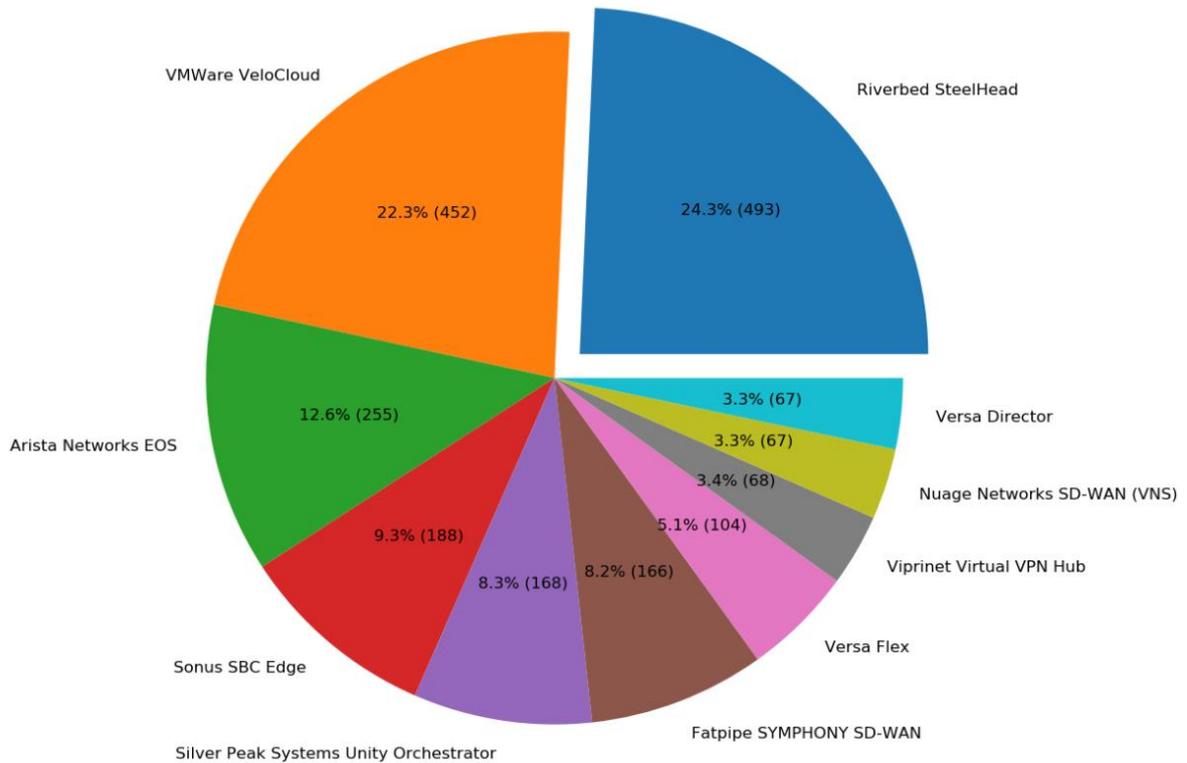

Figure 3. SD-WAN products on the Internet.

## Developed Tools

The developed tools, NMAP scripts, Shodan and Censys search queries are available in the SDWAN-Harvester repository on GitHub [6]. The repository also provides metadata, results and statistics of the Internet scanning implemented on October 2018.

Table 2. SD-WAN versions.

| Product | Latest version | AWS version | Census versions |
|---|---|---|---|
| Cisco vEdge Cloud | 18.3.0 | 17.2.4 | 17.2.0 - 100 % |
| Silver Peak Unity EdgeConnect | 8.1.9.x | 8.1.5.10 | 8.1.9.x - 0.06 %<br>8.1.8.x - 0.03 %<br>8.1.7.x - 45%<br>8.1.6.x - 0.02%<br>8.1.5.10 - 0.04 %<br>< 8.1.5.10 - 54% |
| Citrix NetScaler SD-WAN | 9.3.5 | 9.3.0 | 9.3.1.35 - 54%<br>10.0.0 - 1%<br>10.0.2 - 1% |
| Riverbed SteelConnect Gateway | 2.10 | 2.8.2.16 | - |
| Versa FlexVNF | 16.1R2S1 | - | 100% < 16.1 |
| Arista vEOS Router | 4.21.1F | 4.20.5F | 4.20.5F - 57%<br>< 4.20.5F - 43% |
| VeloCloud Virtual Edge | 2.5.2 | 2.4.1 | - |
| Barracuda CloudGen Firewall | 7.2.2 | 7.2.1 | - |
| Talari Cloud Appliance | R7.3GAP1 | R7.1P1H2 | 7.0 - 30%<br>7.1 - 30%<br>7.2 - 15%<br>7.3 - 23% |

# CONCLUSIONS

We developed active and passive fingerprinting tools for SD-WAN system enumeration. Using these tools we collected information that allows us to evaluate the common security level for SD-WAN products based on gathered statistics on software versions and known vulnerabilities. Most of the SD-WAN systems have known vulnerabilities related to outdated software and insecure configuration.

# ACKNOWLEDGMENTS

The authors would like to thank Nikolay Tkachenko and Max Gorbunov for their help in SD-WAN test bed deployments, fingerprinting, and tools development. Also special thanks to Alexander Sherkin and Anton Desyatov for feedback and comments.